\documentclass[11pt,reqno]{amsart}

\usepackage{amssymb}
\usepackage{url}
\usepackage{hyperref}
\usepackage[dvips]{color}
\usepackage{tikz-cd}
\usepackage[all]{xy}
\SelectTips{cm}{} 

\allowdisplaybreaks[4]

\usepackage{mathrsfs}
\let\mathcal\mathscr
\usepackage[mathscr]{eucal}

\newcommand*{\pd}[2]{\mathchoice{\frac{\partial#1}{\partial#2}}
  {\partial#1/\partial#2}{\partial#1/\partial#2}
  {\partial#1/\partial#2}}

\let\phi=\varphi
\let\kappa=\varkappa
\let\epsilon=\varepsilon

\newcommand*{\abs}[1]{\left|#1\right|}

\newdir{ >}{{}*!/-5pt/\dir{>}}

\theoremstyle{theorem}
\newtheorem{proposition}{Proposition}

\theoremstyle{remark}
\newtheorem{remark}{Remark}

\usepackage{mathrsfs}
\let\mathcal\mathscr
\usepackage[mathscr]{eucal}
\newcommand{\cprime}{\/{\mathsurround=0pt$'$}}

\author{Petr~Voj{\v{c}}{\'{a}}k}
\address{Mathematical Institute, Silesian University in Opava, Na Rybn\'{\i}\v{c}ku 1, 746 01 Opava, Czech Republic}
\email{Petr.Vojcak@math.slu.cz} 
\title[4D MAS equation]{Non-Abelian covering and new recursion operators for the 4D Mart\'inez Alonso - Shabat equation}

\begin{document}

\begin{abstract}
We present new recursion operators for (shadows of nonlocal) symmetries of the 4D Mart\'inez Alonso - Shabat equation $u_{ty} = u_z u_{xy} - u_y u_{xz}$, and we show that their actions can produce new symmetries which are not contained in the Lie algebra of nonlocal symmetries presented in \cite{KrasVoj}. To this end, we construct a non-Abelian covering of the equation in question using the Lax pair with two non-removable parameters.  
\end{abstract}

\maketitle


\section*{Introduction}
\label{sec:introduction}
It is well known that the concept of a (differential) covering plays an important role in the geometry of partial differential equations, see e.g.\ \cite{Boch, Kra2, Kra1}. A covering of a partial differential equation $\mathcal E$ is, roughly speaking, an over-determined system $\mathcal{\tilde E}$ with new additional dependent variables (so-called pseudopotentials or nonlocal variables) such that $\mathcal E$ implies the compatibility conditions of $\mathcal{\tilde E}$. A vector function on an appropriate jet space $J^\infty(\mathcal{\tilde E})$ associated with $\mathcal{\tilde E}$ that depends on finitely many arguments and satisfies linearized version of $\mathcal E$ is called a $\mathcal{\tilde E}$-shadow of $\mathcal E$ associated with the covering $\mathcal{\tilde E}$ \cite{Boch, Kra1, Kra2}. Thus, a $\mathcal{\tilde E}$-shadow can depend also on the nonlocal variables and their derivatives. If a $\mathcal{\tilde E}$-shadow of $\mathcal E$ can be lifted to a full-fledged symmetry of $\mathcal{\tilde E}$ then we call this symmetry a nonlocal symmetry of $\mathcal E$ (see e.g.\ \cite{KrasVoj} for more details).  

In fact, integrable systems in addition to local symmetries usually possess infinite hierarchies of (shadows of) nonlocal symmetries. It can be even argued that precisely such nonlocal hierarchies are the most common feature of known today integrable partial differential systems in any number of independent variables, because (nonlinearizable non-overedetermined) integrable systems in more than two independent variables are generally believed to admit no infinite hierarchies of local symmetries, cf. e.g. \cite{Bla} and references therein.

An important role in the theory of integrable systems is played by recursion operators (ROs) which allow us to generate infinite families of (shadows of nonlocal) symmetries from a suitable seed symmetry. The recursion operators were interpreted in the terms of coverings by Marvan \cite{Mar} (see e.g.\ also \cite{Gut}) as the B\"{a}cklund auto-transformations of the tangent covering of $\mathcal E$. The ROs for multidimensional integrable dispersionless systems are a subject of intense research and there exist several methods for their construction, see \cite{Serg} and references therein for a wider overview of the latter.

The structure of this paper is as follows. In Section \ref{sec:1}, we briefly summarize the previous research of the 4D Mart\'inez Alonso - Shabat equation (shortly the 4D MAS equation) and we specify next problems which we solve here. We also discuss here some reasons for the existence of nonlocal symmetries of the 4D MAS equation which have not been discovered until now. In Section \ref{sec:2}, we introduce new recursion operators for symmetries of the equation in question and we study their actions on (shadows of nonlocal) symmetries. The following Section \ref{sec:3} is devoted to the construction of a non-Abelian covering of the 4D MAS equation and then we present new hierarchies of nonlocal symmetries which arise in this connection.

The new results contained in this paper are very closely related to our previous research on the 4D MAS equation and significantly enhance the earlier results given in \cite{KrasVoj}. For these reasons, we do not repeat here necessary notions and facts from the geometrical theory of PDEs and differential coverings. Instead, we refer reader to the mentioned paper \cite{KrasVoj} or e.g. \cite{Kra1}. We also make an effort to adopt notation from \cite{KrasVoj} as far as it is suitable for our exposition here.

\section{Preliminaries}
\label{sec:1}
To  the best of our knowledge, the 4D MAS equation 
\begin{equation}
  \label{eq:1}
  u_{ty} = u_z u_{xy} - u_y u_{xz} 
\end{equation}
has first appeared in~\cite{MA-Sh-2004} in connection with the study of multi-dimensional systems whose reductions lead to the known $(1+1)$-integrable equations. Nevertheless it also arises, for example, in classification of integrable 4D systems \cite{DFKN}.

It is also a remarkable fact \cite{Mor-Ser, Zhang} that three well-known three-dimensional PDEs, namely the rdDym equation, the universal hierarchy equation and the modified Veronese web equation, can be obtained as reductions of \eqref{eq:1}, cf. e.g.\ \cite{BKMV} and references therein for more details about these equations.

In \cite{Mor}, two coverings (Lax pairs) of \eqref{eq:1} with non-removable parameters were constructed: 
\begin{itemize}
\item one-parameter covering
\begin{equation}
  \label{eq:2}
  \tau_\lambda: \, w_t=u_zw_x-\lambda^{-1}w_z, \quad w_y=\lambda u_yw_x,
\end{equation}
where $0\neq\lambda \in \Bbb R$ and $w$ is the nonlocal variable, \\
\item two-parameter covering
\begin{equation}
  \label{eq:3}
  \tau_{\lambda,\mu}: \, w_t=u_zw_x-(z+\mu)(t+\lambda)^{-1}w_z, \quad w_y=(t+\lambda)(z+\mu)^{-1} u_yw_x,
\end{equation}
where $\lambda,\mu \in \Bbb R$.
\end{itemize}

In \cite{KrasVoj}, we consider a formal expansion $w=\sum_{i\in \Bbb Z}\lambda^iw_i$ and by substituting this expansion into \eqref{eq:2} we have arrived at two different coverings that we call the negative ($\tau^-$) and positive ($\tau^+$) ones. The defining equations for these coverings are
\begin{align}
  \label{eq:4}
  \tau^-\colon\mathcal{E}^-\to\mathcal{E}&\quad
                \left|\begin{array}{l}
                  r_0 = y,\\
                  r_{i,t} = u_zu_y^{-1}r_{i-1,y} - r_{i-1,z},\\[2pt]
                  r_{i,x} = u_y^{-1}r_{i-1,y},\quad i \geq 1,
                \end{array}\right.
  \intertext{and}
  \label{eq:5}
  \tau^+\colon\mathcal{E}^+\to\mathcal{E}&\quad
                \left|\begin{array}{l}
                  q_{-1} = x,\ q_0 = u,\\
                  q_{i,y} = u_y q_{i-1,x},\\
                  q_{i,z} = u_z q_{i-1,x} - q_{i-1,t}, \quad i \geq 1,
                \end{array}\right.
\end{align}
where $r_1$, $r_2\dots$ denote the nonlocal variables in~$\tau^-$ and $q_1$, $q_2,\dots$ are those in~$\tau^+$. Subsequently we have studied the Lie algebra of nonlocal symmetries which arise in the Whitney product $\tau^\pm=\tau^-\oplus \tau^+.$

According to Theorems 1 and 2 in \cite{KrasVoj}, there exist, among others, hierarchies $\{ \Psi_i^j \}_{i \in \Bbb Z}^{j\geq 0},$  $\{ \Omega_i^j \}_{i \in \Bbb Z}^{j\geq 0}$ of nonlocal symmetries of \eqref{eq:1} in $\tau^\pm$. Any symmetry is represented by its generating vector-function, thus, for instance
\begin{align*}
\Psi_i^j\sim\left[\psi_{i,0}^j, \psi_{i,1}^{j,-}, \psi_{i,1}^{j,+}, \dots, \psi_{i,\alpha}^{j,-}, \psi_{i,\alpha}^{j,+},\dots\right],
\end{align*}
where the first component $\psi_{i,0}^j$ is the shadow of $\Psi_i^j$ satisfying the linearized equation 
\begin{equation*}
D_{ty}\psi_{i,0}^j-u_zD_{xy}\psi_{i,0}^j+u_yD_{xz}\psi_{i,0}^j+u_{xz}D_{y}\psi_{i,0}^j-u_{xy}D_{z}\psi_{i,0}^j=0.
\end{equation*} 
Similarly everywhere below $\omega_{i,0}^j$ denotes the shadow of the corresponding nonlocal symmetry~$\Omega_i^j$. 

The direct computations show that there exist four local shadows inside the hierarchies $\{ \Psi_i^j \}_{i \in \Bbb Z}^{j\geq 0},$  $\{ \Omega_i^j \}_{i \in \Bbb Z}^{j\geq 0}$, namely
\begin{equation*}
\psi_{-1,0}^0=-u_z, \quad \psi_{0,0}^0=u_t, \quad \psi_{0,0}^1=tu_t+zu_z, \quad \omega_{0,0}^0=u-zu_z.
\end{equation*}
Let us also note that the lower indices $i\in \Bbb Z$ correspond to the weights of the symmetries in the weight setting
  \begin{equation}
  \label{weights}
    \abs{x} = -1,\qquad \abs{t} = \abs{y} = \abs{u} = 0,\qquad \abs{z} = 1,\qquad \abs{r_i} = -i\qquad \abs{q_i} = i,
  \end{equation}
see \cite{KrasVoj} Remark 3 for more details. On the other hand, upper indices $j \geq 0$ denote levels in the particular hierarchies, see Figure \ref{fig:1} below.

To continue the summary of known structures of the 4D MAS equation, let us also mention the recursion operator which was found in \cite{Mor}. The tangent covering $\mathcal{TE}$ of \eqref{eq:1} is of the form 
\begin{equation}
\label{eq:6}
\begin{aligned}
  u_{ty} &= u_z u_{xy} - u_y u_{xz},\\
  p_{ty} &=u_{xy}p_z - u_{xz}p_y + u_zp_{xy} - u_yp_{xz},
\end{aligned}
\end{equation}
and the B\"{a}cklund auto-transformation that relates two copies of~$\mathcal{TE}$ is
\begin{align}
  \label{eq:7}
  \begin{array}{l}
    D_y(\phi^\prime)=u_yD_x(\phi)-u_{xy}\phi,\\[2pt]
    D_z(\phi^\prime)=-D_t(\phi)+u_zD_x(\phi)-u_{xz}\phi.
  \end{array}
\end{align}

If $\phi$ is a shadow of nonlocal symmetry of Equation~\eqref{eq:1} then~$\phi'$ satisfying \eqref{eq:7} is also a shadow and vice versa. The correspondence $\phi\mapsto\phi'$ defined by relations~\eqref{eq:7} will be denoted by $\mathcal{R}_1^0$ and the opposite one by $\mathcal{R}_{-1}^0$. 

Let us discuss in more details the actions (modulo the images of zero) of both operators on shadows $\psi_{i,0}^j$ of the symmetries from the hierarchy $\{ \Psi_i^j \}_{i \in \Bbb Z}^{j\geq 0}$. The direct computations show that
\begin{equation*}
\mathcal{R}_1^0(\psi_{i,0}^j)=\psi_{i+1,0}^j, \qquad \mathcal{R}_{-1}^0(\psi_{i,0}^j)=\psi_{i-1,0}^j,
\end{equation*}
thus the operator~$\mathcal{R}_1^0$ increases the weights of all shadows $\psi_{i,0}^j$ by~$+1$, while~$\mathcal{R}_{-1}^0$ decreases them by~$-1$. Moreover, neither of them  changes the levels $j$ of the shadows $\psi_{i,0}^j$. So, we can say that these operators act on shadows \textquoteleft horizontally\textquoteright, see Figure \ref{fig:1}. Let us stress that the same diagram is also valid for the shadows $\omega_{i,0}^j$ from the hierarchy $\{ \Omega_i^j \}_{i \in \Bbb Z}^{j\geq 0}$.
\begin{figure}[h]  \centerline{%
  \xymatrix@R+1.15pc@C+1pc{
    &\vdots
    \ar@/^/[d]^-{\mathcal{R}_0^{-1}}
    &\vdots
    \ar@/^/[d]^-{\mathcal{R}_0^{-1}}
    &\vdots
    \ar@/^/[d]^-{\mathcal{R}_0^{-1}}
    &\vdots
    \ar@/^/[d]^-{\mathcal{R}_0^{-1}}
    &\vdots
    \ar@/^/[d]^-{\mathcal{R}_0^{-1}}
\\
    \dots
    \ar@/^/[r]^-{\mathcal{R}_1^0}
    &\psi_{-2,0}^2
    \ar@/^/[l]^-{\mathcal{R}_{-1}^0}
    \ar@/^/[u]^-{\mathcal{R}_0^1}
    \ar@/^/[d]^-{\mathcal{R}_0^{-1}}
    \ar@/^/[r]^-{\mathcal{R}_1^0}
    &\psi_{-1,0}^2
    \ar@/^/[l]^-{\mathcal{R}_{-1}^0}
    \ar@/^/[d]^-{\mathcal{R}_0^{-1}}
    \ar@/^/[u]^-{\mathcal{R}_0^1}
    \ar@/^/[r]^-{\mathcal{R}_1^0}
    &\psi_{0,0}^2
    \ar@/^/[l]^-{\mathcal{R}_{-1}^0}
    \ar@/^/[d]^-{\mathcal{R}_0^{-1}}
    \ar@/^/[u]^-{\mathcal{R}_0^1}
    \ar@/^/[r]^-{\mathcal{R}_1^0}
    &\psi_{1,0}^2
    \ar@/^/[l]^-{\mathcal{R}_{-1}^0}
    \ar@/^/[d]^-{\mathcal{R}_0^{-1}}
    \ar@/^/[u]^-{\mathcal{R}_0^1}
    \ar@/^/[r]^-{\mathcal{R}_1^0}
    &\psi_{2,0}^2
    \ar@/^/[l]^-{\mathcal{R}_{-1}^0}
    \ar@/^/[d]^-{\mathcal{R}_0^{-1}}
    \ar@/^/[u]^-{\mathcal{R}_0^1}
    \ar@/^/[r]^-{\mathcal{R}_1^0}
    &\dots
    \ar@/^/[l]^-{\mathcal{R}_{-1}^0}
\\
    \dots
    \ar@/^/[r]^-{\mathcal{R}_1^0}
    &\psi_{-2,0}^1
    \ar@/^/[u]^-{\mathcal{R}_0^1}
    \ar@/^/[d]^-{\mathcal{R}_0^{-1}}
    \ar@/^/[l]^-{\mathcal{R}_{-1}^0}
    \ar@/^/[r]^-{\mathcal{R}_1^0}
    &\psi_{-1,0}^1
    \ar@/^/[u]^-{\mathcal{R}_0^1}
    \ar@/^/[d]^-{\mathcal{R}_0^{-1}}
    \ar@/^/[l]^-{\mathcal{R}_{-1}^0}
    \ar@/^/[r]^-{\mathcal{R}_1^0}
    &\psi_{0,0}^1
    \ar@/^/[u]^-{\mathcal{R}_0^1}
    \ar@/^/[d]^-{\mathcal{R}_0^{-1}}
    \ar@/^/[l]^-{\mathcal{R}_{-1}^0}
    \ar@/^/[r]^-{\mathcal{R}_1^0}
    &\psi_{1,0}^1
    \ar@/^/[u]^-{\mathcal{R}_0^1}
    \ar@/^/[d]^-{\mathcal{R}_0^{-1}}
    \ar@/^/[l]^-{\mathcal{R}_{-1}^0}
    \ar@/^/[r]^-{\mathcal{R}_1^0}
    &\psi_{2,0}^1
    \ar@/^/[u]^-{\mathcal{R}_0^1}
    \ar@/^/[d]^-{\mathcal{R}_0^{-1}}
    \ar@/^/[l]^-{\mathcal{R}_{-1}^0}
    \ar@/^/[r]^-{\mathcal{R}_1^0}
    &\dots
    \ar@/^/[l]^-{\mathcal{R}_{-1}^0}
\\
    \dots
    \ar@/^/[r]^-{\mathcal{R}_1^0}
    &\psi_{-2,0}^0
    \ar@/^/[u]^-{\mathcal{R}_0^1}
    \ar@/^/[d]^-{\mathcal{R}_0^{-1}}
    \ar@/^/[r]^-{\mathcal{R}_1^0}
    \ar@/^/[l]^-{\mathcal{R}_{-1}^0}
    &\psi_{-1,0}^0
    \ar@/^/[u]^-{\mathcal{R}_0^1}
    \ar@/^/[d]^-{\mathcal{R}_0^{-1}}
    \ar@/^/[r]^-{\mathcal{R}_1^0}
    \ar@/^/[l]^-{\mathcal{R}_{-1}^0}
    &\psi_{0,0}^0
    \ar@/^/[u]^-{\mathcal{R}_0^1}
    \ar@/^/[d]^-{\mathcal{R}_0^{-1}}
    \ar@/^/[r]^-{\mathcal{R}_1^0}
    \ar@/^/[l]^-{\mathcal{R}_{-1}^0}
    &\psi_{1,0}^0
    \ar@/^/[u]^-{\mathcal{R}_0^1}
    \ar@/^/[d]^-{\mathcal{R}_0^{-1}}
    \ar@/^/[r]^-{\mathcal{R}_1^0}
    \ar@/^/[l]^-{\mathcal{R}_{-1}^0}
    &\psi_{2,0}^0
    \ar@/^/[u]^-{\mathcal{R}_0^1}
    \ar@/^/[d]^-{\mathcal{R}_0^{-1}}
    \ar@/^/[r]^-{\mathcal{R}_1^0}
    \ar@/^/[l]^-{\mathcal{R}_{-1}^0}
    &\dots
    \ar@/^/[l]^-{\mathcal{R}_{-1}^0}
 \\
    &?
    \ar@/^/[u]^-{\mathcal{R}_0^1}
    &?
    \ar@/^/[u]^-{\mathcal{R}_0^1}
    &?
    \ar@/^/[u]^-{\mathcal{R}_0^1}
    &?
    \ar@/^/[u]^-{\mathcal{R}_0^1}
    &?
    \ar@/^/[u]^-{\mathcal{R}_0^1}
  }}
  \caption{Action of the recursion operators on shadows $\psi_{i,0}^j$}
  \label{fig:1}
\end{figure}
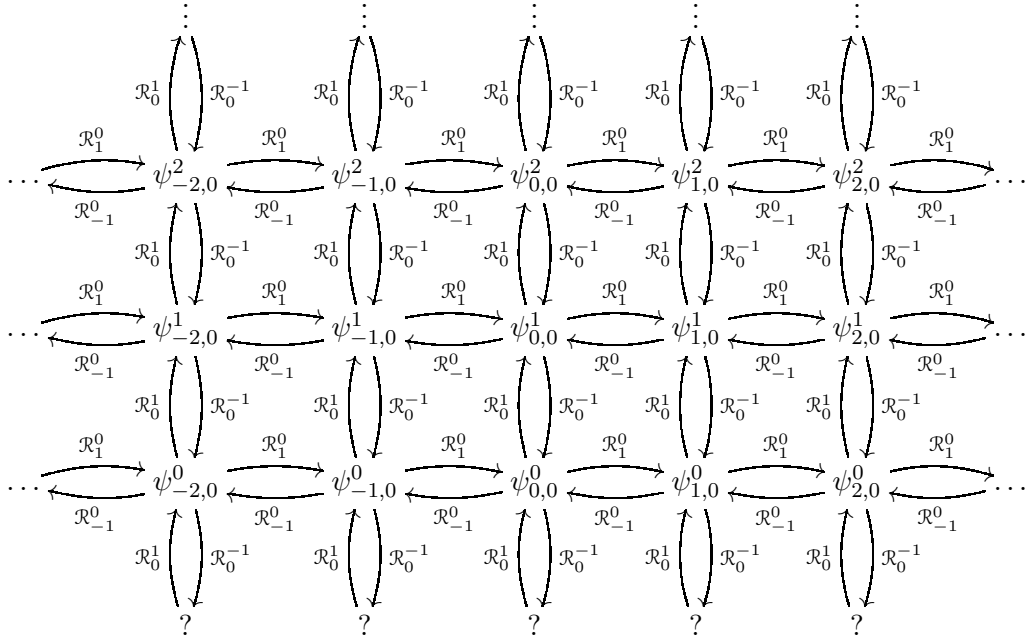

But at this point two natural questions arise:
\begin{itemize}
\item[1)] Is there any other recursion operator $\mathcal{R}_0^1$ which acts on shadows \textquoteleft vertically\textquoteright, i.e. which changes the level $j$ in the hierarchy by $+1$ but does not change the weight $i$?
\item[2)] If the answer to the first question is positive, what will be images of the shadows $\psi_{i,0}^0$ and $\omega_{i,0}^0$ under the action of the inverse operator $\mathcal{R}_0^{-1}$?
\end{itemize}

Mainly the second question indicates that there should be some other nonlocal symmetries of the 4D MAS equation which do not live in the Whitney product $\tau^\pm$. This means that such symmetries need some additional nonlocal variables which cannot be obtained by expansion of the Lax pair \eqref{eq:2}. So, the third natural question is where could one get these new nonlocal variables? We address below all of these problems.

\section{New recursion operators for symmetries}
\label{sec:2}
In this section we answer the first question above. Our procedure of constructing of the desired recursion operators is inspired by the algorithm described in \cite{Kra3}, cf. also \cite{Kra4,Ker}. 

We consider the tangent covering $\mathcal{TE}$ given by \eqref{eq:6} and our first aim is to find differential coverings over $\mathcal{TE}$ linear in the variable $p$ and its derivatives. 
\begin{proposition}
The quantities
\begin{equation}
\label{eq:8}
\theta_1 = \left(T_1 dt + X_1 dx\right) \wedge dy \wedge dz, \qquad \theta_2 = \left(Y_2\,dy + Z_2\,dz\right) \wedge dt \wedge dx,
\end{equation}
where
\begin{align}
  \label{eq:9}
  \begin{array}{ll}
    T_1 = \displaystyle \frac{A(y,z)(u_yp_z-u_zp_y)}{u_y^2},  & X_1 = \displaystyle -\frac{A(y,z)p_y}{u_y^2},\\[4mm]
    Y_2 = B(t,x)(u_yp_x+u_xp_y),\qquad  & Z_2 = B(t,x)(u_zp_x+u_xp_z-p_t),\\[1mm]
  \end{array}
\end{align}
are two-component conservation laws on $\mathcal{TE}$.
Setting A(y,z)=B(t,x)=1, we obtain a covering $\tau_{w_1,w_2}$ of $\mathcal{TE}$ with nonlocal variables $w_1,w_2$, where
\begin{align}
  \label{eq:10}
  \begin{array}{ll}
    w_{1,t} = \displaystyle \frac{u_yp_z-u_zp_y}{u_y^2},  & w_{1,x} = \displaystyle -\frac{p_y}{u_y^2}\\[4mm]
    w_{2,y} = u_yp_x+u_xp_y,\qquad  & w_{2,z} = u_zp_x+u_xp_z-p_t.\\[1mm]
  \end{array}
\end{align}
\end{proposition}
\begin{proof}
This result can be readily verified by straightforward computation.
\end{proof}

In the next step we look for shadows of nonlocal symmetries of Equation \eqref{eq:1} which depend linearly on $p,w_1,w_2$ and their derivatives, i.e. we solve the equation
\begin{equation}
\label{eq:11}
l_{\mathcal{E}}(\bar p)=D_{ty}(\bar p)-u_zD_{xy}(\bar p)+u_yD_{xz}(\bar p)+u_{xz}D_{y}(\bar p)-u_{xy}D_{z}(\bar p)=0
\end{equation}
under the assumption 
\begin{align*}
\bar p=U_0 p + U_1 w_1 + U_2 w_2 &+ U_3 p_t + U_4 p_x + U_5 p_y + U_6 p_z\\
& + U_7 w_{1,y} + U_8 w_{1,z} + U_9 w_{2,t} + U_{10} w_{2,x} + \ldots,
\end{align*}
where $U_i$ are differential functions independent of $p,w_1,w_2$ and their derivatives. Of course, we also always keep in mind that $p$ itself satisfies \eqref{eq:11} together with all differential consequences of this fact. By direct computations we arrive at the following result.
\begin{proposition}
Up to first order jets, there exist six basic shadows of symmetries that depend linearly on the variables $p,w_1,w_2$ and their derivatives, namely
\begin{align}
  \label{eq:12}
  \begin{array}{ll}
\mathcal{R}_0^0: & \bar p_1 = p,\\[2mm]
\mathcal{R}_1^0: & \bar p_2 = -u_x p + w_2,\\[2mm]
\mathcal{R}_{-1}^0: & \bar p_3 = -u_y w_1,\\[2mm]
\mathcal{R}_0^1: & \bar p_4 = t p + zu_yw_1,\\[2mm]
\mathcal{R}_1^1: & \bar p_5 = t(-u_x p + w_2)-zp,\\[2mm]
\mathcal{R}_1^2: & \bar p_6 = t^2(-u_x p + w_2)-2tzp-z^2u_yw_1.
  \end{array}
\end{align}
\end{proposition} 

So, the formulas \eqref{eq:12} provide us six recursion operators $\mathcal{R}_i^j$ for shadows of nonlocal symmetries of \eqref{eq:1}. In all cases, the subscript $i$ indicates the change of the weight of a shadow whereas the upper index $j$ indicates the change of the level in the hierarchies $\{ \Psi_i^j \}_{i \in \Bbb Z}^{j\geq 0},$  $\{ \Omega_i^j \}_{i \in \Bbb Z}^{j\geq 0}$.

Let us briefly discuss the particular operators.
\begin{itemize}
\item[a)] $\mathcal{R}_0^0$ is obviously the identical operator and thus is of no interest.\\[-3mm]
\item[b)] $\mathcal{R}_1^0$ corresponds to Morozov's recursion operator \eqref{eq:7} and $\mathcal{R}_{-1}^0$ is its inverse.\\[-3mm]
\item[c)] $\mathcal{R}_0^1$ is the desired \textquoteleft vertical\textquoteright \;operator for our diagram in Figure \ref{fig:1} above and as far as we know it has not been published in literature yet. Eliminating the nonlocal variable $w_1$, we obtain its conventional form  
\begin{equation}
\label{eq:13}
\begin{aligned}
    u_yD_x(\phi^\prime)-u_{xy}\phi^\prime &= t(u_yD_x(\phi)-u_{xy}\phi)-zD_y(\phi),\\[2pt]
    -D_t(\phi^\prime)+u_zD_x(\phi^\prime)-u_{xz}\phi^\prime &= t(-D_t(\phi)+u_zD_x(\phi)-u_{xz}\phi)-zD_z(\phi)-\phi.
\end{aligned} 
\end{equation}
Similarly as in \eqref{eq:7}, $\mathcal{R}_0^1$ means the correspondence $\phi \mapsto \phi^\prime$ defined by \eqref{eq:13}. The opposite one will be denoted by $\mathcal{R}_0^{-1}$.\\[-3mm]
\item[d)] $\mathcal{R}_1^1$ is a composition of $\mathcal{R}_0^1$ and $\mathcal{R}_1^0$, thus from this point of view it is not too interesting. On the other hand, its conventional form
\begin{equation*}
\begin{aligned}
    D_y(\phi^\prime) &= t(u_yD_x(\phi)-u_{xy}\phi)-zD_y(\phi),\\[2pt]
    D_z(\phi^\prime) &= t(-D_t(\phi)+u_zD_x(\phi)-u_{xz}\phi)-zD_z(\phi)-\phi
\end{aligned} 
\end{equation*}
is simpler than \eqref{eq:13}, so one can use it for easier crossing between levels.\\[-3mm]
\item[e)] $\mathcal{R}_1^2 = \mathcal{R}_0^1 \circ \mathcal{R}_1^1 = \mathcal{R}_0^1 \circ \mathcal{R}_0^1 \circ \mathcal{R}_1^0$ and thus we can more or less omit it. 
\end{itemize}

\begin{remark}
One can also simply derive from \eqref{eq:12} the following relations:
\begin{align}
  \label{eq:14}
  \begin{array}{ll}
\mathcal{R}_0^1 = t\mathcal{R}_0^0-z\mathcal{R}_{-1}^0,\\[2mm]
\mathcal{R}_1^1= t\mathcal{R}_1^0-z\mathcal{R}_0^0,\\[2mm]
\mathcal{R}_1^2 = t\mathcal{R}_1^1-z\mathcal{R}_0^1 = t^2\mathcal{R}_1^0-2tz\mathcal{R}_0^0+z^2\mathcal{R}_{-1}^0.
  \end{array}
\end{align}

We conjecture that it is possible to construct in this way infinitely many recursion operators for shadows of nonlocal symmetries of \eqref{eq:1}. More precisely, we claim that for each couple $\mathcal{R}_i^j$, $\mathcal{R}_{i-1}^j$ of recursion operators, where $i \in \Bbb Z$ and $j\geq0$ we obtain a new recursion operator $\mathcal{R}_{i}^{j+1}$ of the form
$$\mathcal{R}_{i}^{j+1}=t\mathcal{R}_i^j-z\mathcal{R}_{i-1}^j =  \sum_{m=0}^j(-1)^m\binom{j}{m}t^{j-m}z^m\mathcal{R}_{i-m}^0,$$ 
where $\mathcal{R}_i^0 = \underbrace{\mathcal{R}_1^0 \circ \ldots \circ \mathcal{R}_1^0}_{i-times}$ for $i\geq 1$, and $\mathcal{R}_i^0 = \underbrace{\mathcal{R}_{-1}^0 \circ \ldots \circ \mathcal{R}_{-1}^0}_{-i-times}$ for $i\leq -1$. 

Nevertheless, we also claim that actions of such recursion operators $\mathcal{R}_i^{j+1}$ will always be expressible as a composition of actions of $\mathcal{R}_0^1$ and $\mathcal{R}_1^0$ or $\mathcal{R}_{-1}^0$. From this point of view, the recursion operators $\mathcal{R}_0^1$ and $\mathcal{R}_1^0$ and their inversions are much more essential than the other ones and we will call them the basic recursion operators.

Finally, let us note that all our presented conjectures are also supported by the formulas (14) and (21) in \cite{KrasVoj} and Theorems~1 and 2 ibid.
\end{remark}

\begin{remark}
In the Whitney product $\tau^\pm$, Equation \eqref{eq:1} also admits two hierarchies of nonlocal symmetries $\{ \Xi_i(A)\}_{i \in \Bbb Z}$, $\{ \Upsilon_j (B)\}_{j \in \Bbb Z}$ which allow arbitrary functions $A=A(y,z)$ and $B=B(t,x)$, see \cite{KrasVoj} for more details. However, we postpone the exposition of the actions of the recursion operator $\mathcal{R}_0^1$ on the corresponding shadows to Section \ref{sec:4}. \\
\end{remark}

\section{Non-Abelian covering and new hierarchies of nonlocal symmetries}
\label{sec:3}
As one can readily verify, the correspondence $\phi \mapsto \phi^\prime$ given by \eqref{eq:13} works perfectly for all the shadows we have known so far. On the other hand, we can quickly get into troubles with the opposite correspondence $\phi^\prime \mapsto \phi$, i.e. when we apply the recursion operator $\mathcal{R}_0^{-1}$ (cf. discussion at the end of Section \ref{sec:1}). However, it turns out that the two-parameter Lax pair \eqref{eq:3} allows us to overcome these problems. 

Consider an expansion $w=\sum_{i \in \Bbb Z}\mu^iw_i$ and substitute it into \eqref{eq:3}. Setting $\lambda=0$, we get an infinite-dimensional covering over \eqref{eq:1} defined by
\begin{equation}
\label{eq:15}
w_{i-1,y} = tu_yw_{i,x}-zw_{i,y}, \qquad w_{i-1,z}=t(u_zw_{i,x}-w_{i,t})-zw_{i,z}, \qquad i \in \Bbb Z.
\end{equation}
Assuming $w_i=0$ for $i<0$, we obtain the covering
\begin{align}
  \label{eq:16}
  \tau_w\colon\mathcal{E}^w\to\mathcal{E}&\quad
                \left|\begin{array}{l}
                   w_{i,y} = z^{-1}(tu_yw_{i,x} - w_{i-1,y}),\\[2pt]
                   w_{i,z} = z^{-1}(tu_zw_{i,x}-tw_{i,t} - w_{i-1,z}),\quad i \geq 0,
                \end{array}\right.
\end{align}
which is non-Abelian, i.e.\ the nonlocal variables $w_i$ are not determined by conservation laws of the underlying equation (for other examples of non-Abelian coverings see e.g.\ \cite{BKMV, Kis, Kra1}).

Having the new nonlocal variables $w_i$ for our disposal, we can extend the Whitney product $\tau^\pm$ to a bigger covering $\tau_w^\pm=\tau_w \oplus \tau^\pm$. We show below that all remaining problems can be solved just in this extended covering $\tau_w^\pm$.

Concerning an assignment of weights to the new nonlocal variables $w_i$, there is a lot of freedom how to do it. Indeed, it follows directly from \eqref{weights} and \eqref{eq:16} that the weight of the variable $w_0$ can be set completely arbitrarily and further we have
\begin{align}
  \label{weights_wi}
   |w_{i}| = |w_{i-1}|-1, \quad i \geq 1.
\end{align}
In what follows we put $|w_0|=0$ and thus $|w_i|=-i$ for all $i$. 

\begin{remark}
Note that the opposite reduction of \eqref{eq:15}, that is $w_i=0$ for $i>0$, produces a covering that is equivalent to the covering $\tau^+$ given by \eqref{eq:5}. One can also expand the Lax pair \eqref{eq:3} in parameter $\lambda$ and set $\mu=0$. In this case, we again obtain the covering \eqref{eq:16} and a covering that is equivalent to the covering $\tau^-$ determined by \eqref{eq:4}. In this sense, it would be possible to derive all necessary nonlocal variables using the two-parameter Lax pair \eqref{eq:3} only.
\end{remark}

The action of the recursion operator $\mathcal{R}_0^{-1}$ become closed in the covering $\tau_w^\pm$ and, moreover, it produces infinite series of new nonlocal shadows $\psi_{i,0}^{j-1}=\mathcal{R}_0^{-1}(\psi_{i,0}^j)$ and $\omega_{i,0}^{j-1}=\mathcal{R}_0^{-1}(\omega_{i,0}^j)$, where $j\leq0$. As a demonstration, we present here the exact forms of $\psi_{-1,0}^{-1}$, $\psi_{0,0}^{-1}$ and $\omega_{0,0}^{-1}$ which appear as images of local shadows:
\begin{align*}
  \begin{array}{lll}
\psi_{-1,0}^0 =  -u_z & \mapsto &
\psi_{-1,0}^{-1} = \mathcal{R}_0^{-1}(\psi_{-1,0}^0) = \displaystyle -\frac{w_{0,t}}{tw_{0,x}}-\frac{zw_1}{t^2w_{0,x}},\\[5mm]
\psi_{0,0}^0 = u_t & \mapsto & 
\psi_{0,0}^{-1} = \mathcal{R}_0^{-1}(\psi_{0,0}^0) = \displaystyle \frac{u_t}{t}-\frac{zw_{0,t}}{t^2w_{0,x}}-\frac{z^2w_1}{t^3w_{0,x}},\\[5mm]
\omega_{0,0}^0 = u-zu_z & \mapsto & 
\omega_{0,0}^{-1} = \mathcal{R}_0^{-1}(\omega_{0,0}^0) = \displaystyle \frac{u}{t}-\frac{zw_{0,t}}{tw_{0,x}}.
  \end{array}
\end{align*}

It remains to show that the new shadows $\psi_{i,0}^j$ and $\omega_{i,0}^j$, $j<0$, can be lifted to the full-fledged nonlocal symmetries
\begin{align*}
\Psi_i^j\sim\left[\psi_{i,0}^j, \psi_{i,0}^{j,w}, \psi_{i,1}^{j,-}, \psi_{i,1}^{j,+}, \psi_{i,1}^{j,w}, \dots, \psi_{i,\alpha}^{j,-}, \psi_{i,\alpha}^{j,+}, \psi_{i,\alpha}^{j,w},\dots\right],
 \intertext{and}
\Omega_i^j\sim\left[\omega_{i,0}^j, \omega_{i,0}^{j,w}, \omega_{i,1}^{j,-}, \omega_{i,1}^{j,+}, \omega_{i,1}^{j,w}, \dots, \omega_{i,\alpha}^{j,-}, \omega_{i,\alpha}^{j,+}, \omega_{i,\alpha}^{j,w},\dots\right] 
\end{align*}
in the Whitney product $\tau_w^{\pm}$. We prove the latter in the following proposition. 
\begin{proposition}
\label{prop:3}
The shadows $\psi_{i,0}^j$, and $\omega_{i,0}^{j-1}$, $i\in \Bbb Z, j\leq 0$, admit lifts to the covering $\tau_w^{\pm}$, i.e.\ there exist the hierarchies of full-fledged nonlocal symmetries $\{ \Psi_i^j \}_{i\in \Bbb Z}^{j <0}$ and $\{ \Omega_i^j \}_{i\in \Bbb Z}^{j <0}$ of \eqref{eq:1} in~$\tau_w^{\pm}$.
\end{proposition}

\begin{proof}
The lifts of the shadows $\psi_{0,0}^0$, $\psi_{0,0}^{-1}$ and $\omega_{0,0}^{-1}$ are given explicitly as follows:
\begin{itemize}
\item for the shadow $\psi_{0,0}^0$
\begin{align*}
\psi_{0,\alpha}^{0,w} &= t^{-1}((\alpha+1) z w_{\alpha+1}+\alpha w_{\alpha})+w_{\alpha,t}, \quad \alpha\geq 0, &\\
\psi_{0,\kappa}^{0,-} &= r_{\kappa,t}, \quad \psi_{0,\kappa}^{0,+} = q_{\kappa,t}, \quad \kappa \geq 1,
\end{align*}

\item for the shadow $\psi_{0,0}^{-1}$
\begin{align*}
\psi_{0,\alpha}^{-1,w} &= t^{-2}\left( z^2 \left((\alpha+2)w_{\alpha+2}-\frac{w_1w_{\alpha+1,x}}{w_{0,x}}\right)+2(\alpha+1)zw_{\alpha+1}+\alpha w_\alpha\right)\\
&+ t^{-1}\left(z\left(w_{\alpha+1,t}-\frac{w_{0,t}w_{\alpha+1,x}}{w_{0,x}}\right) +  w_{\alpha,t}\right), \quad \alpha\geq 0, &\\[3mm]
\psi_{0,1}^{-1,-} &= r_{1,x} \left( \frac{w_{0,t}}{tw_{0,x}}+\frac{zw_1}{t^2w_{0,x}} \right),\\[2mm]
\psi_{0,\kappa}^{-1,-} &= r_{\kappa-1,y}\psi_{0,1}^{-1,-}+z^{-1}\left(t\psi_{0,\kappa-1}^{-1,-}-r_{1,t}r_{\kappa-2,y}+ r_{\kappa-2,z}\right),\\[3mm]
\psi_{0,1}^{-1,+} &= \frac{q_{1,t}}{t}+t^{-2}z\left(u_t-\frac{u_xw_{0,t}}{w_{0,x}}\right)-\frac{z^2(u_xw_1+w_{0,t})}{t^3w_{0,x}}-\frac{z^3w_1}{t^4w_{0,x}},\\[2mm]
\psi_{0,\kappa}^{-1,+} &= t^{-1}\left(z\psi_{0,\kappa-1}^{-1,+}+q_{\kappa,t}\right)-q_{\kappa-1,x}\left( \frac{zw_{0,t}}{t^2w_{0,x}} + \frac{z^2w_1}{t^3w_{0,x}} \right), \quad \kappa \geq 2,
\end{align*}

\item for the shadow $\omega_{0,0}^{-1}$
\begin{align*}
\omega_{0,\alpha}^{-1,w} &= z\left( w_{\alpha+1,t} -\frac{(\alpha+1)w_{\alpha+1}}{t}-\frac{w_{0,t}w_{\alpha+1,x}}{w_{0,x}}\right)-\frac{\alpha w_\alpha}{t}, \quad \alpha\geq 0, &\\[3mm]
\omega_{0,1}^{-1,-} &= \frac{r_{1,x} w_{0,t}}{w_{0,x}}-r_{1,t},\\[2mm]
\omega_{0,\kappa}^{-1,-} &=z^{-1}\left( t \omega_{0,\kappa-1}^{-1,-} +(\kappa-1)r_{\kappa-1}\right) + r_{\kappa-1,y} \omega_{0,1}^{-1,-} + r_{\kappa-1,z},\\[3mm]
\omega_{0,1}^{-1,+} &=t^{-1} \left( 2q_1+zu_t -\frac{zu_xw_{0,t}}{w_{0,x}}\right) + t^{-2} \left( zu-\frac{z^2w_{0,t}}{w_{0,x}} \right),\\[2mm]
\omega_{0,\kappa}^{-1,+} &= t^{-1}\left(z \left( \omega_{0,\kappa-1}^{-1,+} + q_{\kappa-1,t} - \frac{q_{\kappa-1,x}w_{0,t}}{w_{0,x}}\right) + (\kappa+1)q_\kappa \right), \quad \kappa \geq 2.
\end{align*}
\end{itemize}

The next step consists in the construction of two basic infinite hierarchies of nonlocal symmetries $\{ \Psi_0^j \}^{j <0}$ and $\{ \Omega_0^j \}^{j <0}$. As one can verify by tedious computations, it holds
\begin{equation*}
\psi_{0,0}^j = \frac{1}{j+1}\{\Psi_0^{j+1}, \Psi_0^0\}_1, \quad \omega_{0,0}^j = \frac{1}{j+1}\{\Omega_0^{j+1},  \Psi_0^0\}_1, \quad j\leq-2,
\end{equation*} 
where the lower index 1 means the first component of the corresponding Jacobi bracket. Thus, the required components of the lifts can be obtained from these Jacobi brackets.

Finally, following the technique of the proof of Theorem 1 in \cite{KrasVoj}, it is possible to prove the formulas
\begin{equation}
\label{eq:17}
\Psi_i^j=t\Psi_{i}^{j-1}-z\Psi_{i-1}^{j-1},\qquad \Omega_i^j=t\Omega_{i}^{j-1}-z\Omega_{i-1}^{j-1}, \qquad j \in \Bbb Z.
\end{equation}
This means that the basic hierarchies $\{ \Psi_0^j \}^{j <0}$ and $\{ \Omega_0^j \}^{j <0}$ allow us to get each other symmetry recursively from \eqref{eq:17}.
\end{proof}

\section{Nonlocal symmetries involving arbitrary functions}
\label{sec:4}
In this section we will study hierarchies of nonlocal symmetries of the equation \eqref{eq:1} which admit arbitrary functions. The complete description of two such hierarchies, namely $\{ \Xi_i(A)\}_{i \in \Bbb Z}$ and $\{ \Upsilon_j(B)\}_{j \in \Bbb Z}$, where $A=A(y,z)$ and $B=B(t,x)$, is given in \cite{KrasVoj}. Let us briefly summarize some basic facts about members of these hierarchies:
\begin{itemize}
\item All nonlocal symmetries $\Xi_i(A)$, $i \geq 1$, and $\Upsilon_j(B)$, $j \leq -2$ are so-called invisible symmetries, i.e.\ their shadows are trivially zeros.
\item The weights of the nonlocal symmetries $\Xi_i(A)$, $i \leq 0$, and $\Upsilon_j(B)$, $j \geq -1$ are assigned assuming $A=y$ and $B=x$, respectively.
\item The symmetries $\Xi_0(A)$, $\Upsilon_{-1}(B)$ and $\Upsilon_0(B)$ are the only ones with nontrivial local shadows that are
\begin{equation*}
\xi_{0,0}(A)=-Au_y, \quad \upsilon_{-1,0}(B) = B, \quad \upsilon_{0,0}(B)=-Bu_x-zB_t+uB_x.
\end{equation*}
\item The recursion operator $\mathcal R_1^0$ and its inverse $\mathcal R_{-1}^0$ act within these hierarchies and, until the image or pre-image is nonzero, they do not change either function $A$ or function $B$, see Figure \ref{fig:2} below. Recall that all actions of the recursion operators are permanently computed modulo images of zero.
\end{itemize}

According the last point, there are no troubles with the operator $\mathcal R_1^0$ and its inversion. Nevertheless, the situation becomes different if we study the actions of the operators $\mathcal R_0^1$ and $\mathcal R_0^{-1}$. Then it turns out that we need some new nonlocal shadow which will serve us as a pre-image of zero with respect to $\mathcal R_0^1$. However, the direct computation shows that such shadow can be obtained in the covering $\tau_w^\pm$ and the exact formula for it is
\begin{equation}
\label{eq:18}
\theta_0^0(W) = \frac{W}{tw_{0,x}},
\end{equation} 
where $W=W(w_0,t^{-1}z)$ is an arbitrary function.

\begin{proposition}
The shadow $\theta_0^0(W)$ can be lifted to the full-fledged nonlocal symmetry
\begin{align*}
\Theta^0(W)\sim\left[\theta_0^0(W), \theta_0^{0,w}(W), \theta_1^{0,-}(W), \theta_1^{0,+}(W), \theta_1^{0,w}(W), \dots, \theta_\alpha^{0,-}(W), \theta_\alpha^{0,+}(W), \theta_\alpha^{0,w}(W),\dots\right]
\end{align*}
in the Whitney product $\tau_w^{\pm}$. 
\end{proposition} 
\begin{proof}
The components $\theta_\alpha^{0,-}(W)$ and $\theta_\alpha^{0,+}(W)$ of the lift are given explicitly as follows:
\begin{align*}
\theta_\alpha^{0,-}(W) &= -\frac{W}{z^\alpha u_y w_{0,x}} \sum \limits_{i=0}^{\alpha-1}z^i t^{\alpha-1-i}r_{i,y},\\[2mm]
\theta_\alpha^{0,+}(W) &= \frac{W}{t^{\alpha+1} w_{0,x}} \sum \limits_{i=0}^{\alpha}z^{\alpha-i} t^iq_{i-1,x}, \quad \alpha \geq 1.
\end{align*}

To obtain a concise form of the components $\theta_\alpha^{0,w}(W)$, let us introduce the operator 
$$\mathcal{Z} = \pd{}{z}+\sum \limits_{i=0}^\infty (i+1)w_{i+1}\pd{}{w_i}$$
and the quantities $R_\alpha(W)$, $\alpha=0,1,2,\ldots,$ defined by induction in the following way:
$$R_0(W)=W, \qquad R_\alpha(W)=\frac{1}{\alpha}\mathcal{Z} (R_{\alpha-1}(W)), \quad \alpha \geq 1.$$
Then one can verify that
\begin{align*}
\theta_\alpha^{0,w}(W) &= \frac{W w_{\alpha+1,x}}{w_{0,x}}-R_{\alpha+1}, \quad \alpha \geq 0.
\end{align*}
\end{proof}

Now using the symmetry $\Psi_0^0$, we can even construct  an infinite hierarchy $\{ \Theta^j (W)\}_{j\geq 0}$ of nonlocal symmetries which admit arbitrary functions $W=W(w_0,t^{-1}z)$. The members of this hierarchy are defined recursively by the Jacobi brackets as follows:
$$\Theta^j(W) =\frac{1}{j} \{ \Psi_0^0, \Theta^{j-1}(W) \}, \quad j\geq 1.$$
\begin{remark}
As all members of the hierarchy $\{ \Theta^j (W)\}_{j\geq 0}$ have the same weight (which depends on the particular choice of the function $W$) we use only superscripts $j$'s when naming them. For example, if we put $W=t^{-1}z$ then $|\Theta^j (W)|=0$ for all $j$.
\end{remark}

The actions of all basic recursion operators on the corresponding shadows are illustrated in the next diagram. The modifications of the functions, if any, under the particular actions are described in the brackets. For instance, $(A \rightarrow -zA)$ means that the function $A$ is modified to the function $-zA$, etc. 

\begin{figure}[h]
\begin{center}
\begin{tikzcd}[row sep=2cm, column sep=1cm]
\dots
\arrow[bend left=10, "\mathcal R_1^0"{anchor=south}]{r}
\arrow[bend left=70, "\mathcal R_0^{-1}\, (A \rightarrow -A/z)"{anchor=south}]{r}  
&
\xi_{-1,0}(A)
\arrow[bend left=10, "\mathcal R_1^0"{anchor=south}]{r} 
\arrow[bend left=10, "\mathcal R_{-1}^0"{anchor=north}]{l} 
\arrow[bend left=60, "\mathcal R_0^{-1}\, (A \rightarrow -A/z)"{anchor=south}]{r}
\arrow[bend left=70, "\mathcal R_0^1\, (A \rightarrow -zA)"{anchor=north}]{l}  
&
\xi_{0,0}(A)
\arrow[bend left=10, "\mathcal R_1^0"{anchor=south}]{r} 
\arrow[bend left=10, "\mathcal R_{-1}^0"{anchor=north}]{l} 
\arrow[bend left=60, "\mathcal R_0^{-1}"{anchor=south}]{r} 
\arrow[bend left=60, "\mathcal R_0^1\, (A \rightarrow -zA)"{anchor=north}]{l} 
&
0 
\arrow[bend left=10, "\mathcal R_1^0"{anchor=south}]{r} 
\arrow[bend left=10, "\mathcal R_{-1}^0"{anchor=north}]{l} 
\arrow[bend left=10, "\mathcal R_0^{-1}"{anchor=west}]{d} 
\arrow[bend left=60, "\mathcal R_0^1"{anchor=north}]{l} 
& 
\upsilon_{-1,0}(B) 
\arrow[bend left=10, "\mathcal R_1^0"{anchor=south}]{r} 
\arrow[bend left=10, "\mathcal R_{-1}^0"{anchor=north}]{l}  
\arrow[ "\mathcal{R}_0^{-1}\, (B \rightarrow B/t)"{anchor=south}, out=140, in=40, loop, distance=2.0cm]
\arrow[ "\mathcal{R}_0^1\, (B \rightarrow tB)"{anchor=north}, out=320, in=220, loop, distance=2.0cm]
&  
\upsilon_{0,0}(B)
\arrow[bend left=10, "\mathcal R_1^0"{anchor=south}]{r} 
\arrow[bend left=10, "\mathcal R_{-1}^0"{anchor=north}]{l} 
\arrow[ "\mathcal{R}_0^{-1}\, (B \rightarrow B/t)"{anchor=south}, out=140, in=40, loop, distance=2.0cm]
\arrow[ "\mathcal{R}_0^1\, (B \rightarrow tB)"{anchor=north}, out=320, in=220, loop, distance=2.0cm]
&
\dots
\arrow[bend left=10, "\mathcal R_{-1}^0"{anchor=north}]{l} 
\\ 
&
&
&
\theta_0^0(W)
\arrow[bend left=10, "\mathcal R_0^1"{anchor=east}]{u}
\arrow[ "\mathcal{R}_{-1}^0\, (W \rightarrow tW/z)"{anchor=east}, out=230, in=130, loop, distance=2.0cm] 
\arrow[ "\mathcal{R}_1^0\, (W \rightarrow zW/t)"{anchor=west}, out=50, in=310, loop, distance=2.0cm]
\arrow[bend right=30, "\mathcal R_0^{-1}"{anchor=west}]{dr}
\\
&
&
&
&
\theta_0^1(W)
\arrow[bend left=10, "\mathcal R_0^{-1}"{anchor=south}]{r} 
\arrow[ "\mathcal{R}_1^0\, (W \rightarrow zW/t)"{anchor=south}, out=140, in=40, loop, distance=2.0cm]
\arrow[ "\mathcal{R}_{-1}^0\, (W \rightarrow tW/z)"{anchor=north}, out=320, in=220, loop, distance=2.0cm]
\arrow[bend left=55, "\mathcal R_0^1"{anchor=east}]{ul}
&
\theta_0^2(W)
\arrow[bend left=10, "\mathcal R_0^{-1}"{anchor=south}]{r} 
\arrow[bend left=10, "\mathcal R_0^1"{anchor=north}]{l} 
\arrow[ "\mathcal{R}_1^0\, (W \rightarrow zW/t)"{anchor=south}, out=140, in=40, loop, distance=2.0cm]
\arrow[ "\mathcal{R}_{-1}^0\, (W \rightarrow tW/z)"{anchor=north}, out=320, in=220, loop, distance=2.0cm]
&
\dots
\arrow[bend left=10, "\mathcal R_0^1"{anchor=north}]{l} 
\end{tikzcd}
\end{center}
  \caption{Action of the basic recursion operators on shadows involving arbitrary functions}
  \label{fig:2}
\end{figure}
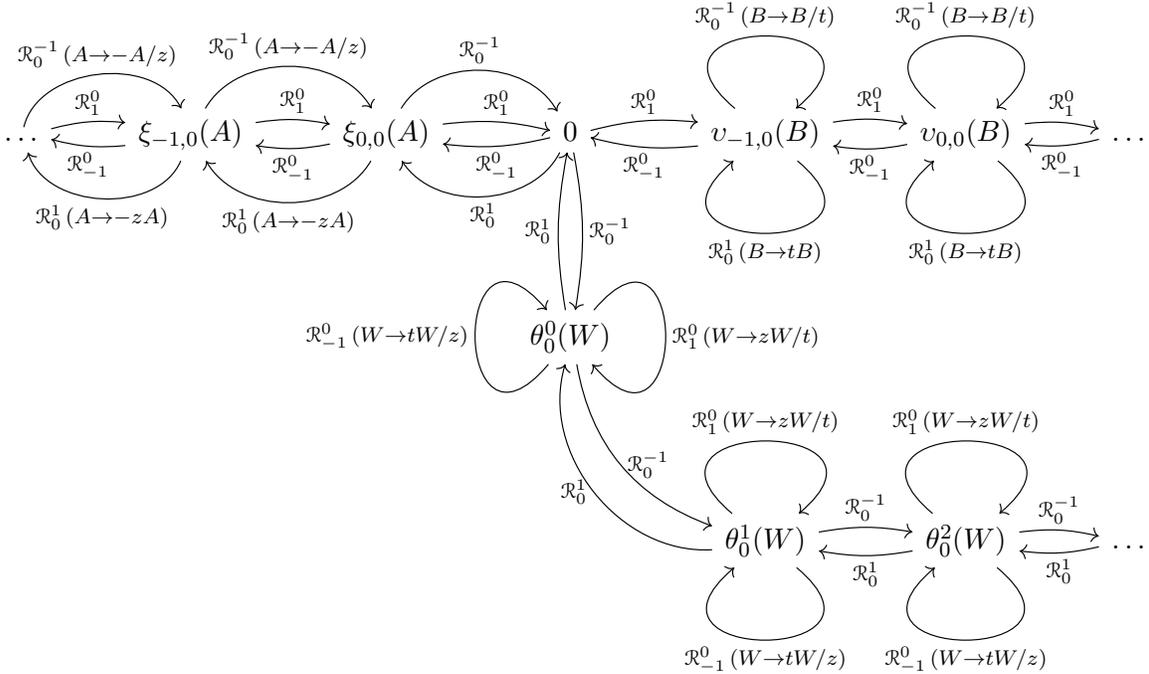
As a final observation, let us note that Figures \ref{fig:1} and \ref{fig:2} indicate a commutativity of the basic recursion operators. Of course, this fact can be also verified by the straightforward computations.  

\section{Conclusions}
It follows from our exposition above that the basic recursion operators $\mathcal R_1^0$, $\mathcal R_0^1$ and their inversions $\mathcal R_{-1}^0$ and $\mathcal R_0^{-1}$ allow us to obtain all shadows of nonlocal symmetries of the 4D MAS equation \eqref{eq:1} in the covering $\tau_w^\pm$ from a small amount of known shadows. Indeed, if we take, for example, the trivial shadow and the local shadows $\psi_{0,0}^0=u_t$ and $\omega_{0,0}^0 = u-zu_z$ as the seeds then each other shadow can be obtained by the particular sequential application of the basic recursion operators to one of them.

An extremely interesting feature of the newly presented recursion operators for the 4D MAS equation \eqref{eq:1} is their explicit dependence on independent variables, see \eqref{eq:13} and Remark 1 above. This generally seems to be a very unusual property among the known today recursion operators for equations in more than two independent variables, especially in the situation when the equations do not depend explicitly on the independent variables. In fact, we have been able to find \cite{Krug} only four known examples of such recursion operators, however, in all cases the corresponding equation itself depends explicitly on the independent variables too. 

We are almost sure that there are also other four-dimensional integrable linearly degenerate equations (in the sense of \cite{Fer}) which admit a pair of recursion operators for shadows of nonlocal symmetries, where one of them acts \textquoteleft horizontally\textquoteright\ and the second one \textquoteleft vertically\textquoteright\ (in analogy with Figure \ref{fig:1} above), and whose Lie algebras of nonlocal symmetries evince the same or very similar structure as in the case of the 4D MAS equation. In our opinion, such equation should be, for example, the modified Mart\'inez Alonso-Shabat equation \cite{Mor-Ser}
\begin{equation*}
u_y u_{xz} + \alpha u_xu_{ty} - (u_z+\alpha u_t)u_{xy} = 0, \quad \alpha \neq 0,
\end{equation*} 
whose \textquoteleft horizontal\textquoteright\ recursion operator was recently found in \cite{Bar}. It would also be interesting to examine in this sense the 4D equations presented in \cite{DFKN}, especially the so-called 4D Pavlov equation
\begin{equation*}
u_{yz}=u_{tx}+u_xu_{xy}-u_yu_{xx},
\end{equation*}  
which can be obtained, like the 4D MAS equation, as a reduction of the five-dimensional equation
\begin{equation}
\label{eq:21}
u_{yz}=u_{ts}+u_su_{xz}-u_zu_{xs},
\end{equation}  
see \cite{BKMV2} and references therein for further details. And subsequently it could be very instructive to study the five-dimensional equation \eqref{eq:21} itself. We intend to deal with these issues in the forthcoming research.

\section*{Acknowledgments}
\label{sec:acknowledgments}

The symbolic computations were performed using the software \textsc{Jets} \cite{Jets}. Computational resources were supplied by the project "e-Infrastruktura CZ" (e-INFRA CZ LM2018140 ) supported by the Ministry of Education, Youth and Sports of the Czech Republic. The research was supported by the Ministry of Education, Youth and Sports of the Czech Republic (MSMT CR) under RVO funding for IC47813059.

The author is also grateful to A. Sergyeyev for his valuable recommendations.

\end{document}